# An Intelligent Trust Cloud Management Method for Secure Clustering in 5G enabled Internet of Medical Things

Liu Yang, Keping Yu, *Member*, *IEEE*, Simon X. Yang, Senior *Member*, *IEEE*,

Chinmay Chakraborty, *Member*, *IEEE*, Yinzhi Lu, and Tan Guo

*Abstract*—5G edge computing enabled Internet of Medical Things (IoMT) is an efficient technology to provide decentralized medical services while Device-to-device (D2D) communication is a promising paradigm for future 5G networks. To assure secure and reliable communication in 5G edge computing and D2D enabled IoMT systems, this paper presents an intelligent trust cloud management method. Firstly, an active training mechanism is proposed to construct the standard trust clouds. Secondly, individual trust clouds of the IoMT devices can be established through fuzzy trust inferring and recommending. Thirdly, a trust classification scheme is proposed to determine whether an IoMT device is malicious. Finally, a trust cloud update mechanism is presented to make the proposed trust management method adaptive and intelligent under an open wireless medium. Simulation results demonstrate that the proposed method can effectively address the trust uncertainty issue and improve the detection accuracy of malicious devices.

## I. INTRODUCTION

WITH the rapid expansion in the deployment of Internet of Things (IoT) devices and increasing desire to make healthcare more cost effective, proactive, and personalized, Internet of Medical Things (IoMT) is a promising paradigm that serves all aspects of the medical field [1]-[4]. Although IoMT can provide comprehensive healthcare monitoring services, the excessive requirements still overload the medical center while hindering the development of IoMT. 5G edge computing enabled IoMT is considered as a favorable technique to eliminate such an obstacle [5]. The medical analysis task is offloaded to the edge server in proximity while the delay constraint of the time-sensitive tasks can be satisfied. 5G edge computing enabled IoMT system can be divided into two sub-networks, one is Wireless Body Area Network (WBAN), and the other is 5G cellular network. WBAN is the network consists of various tiny sensors with limited storage, power, and computing resources [6]. The sensors are implanted either on or inside the body of the patient to collect personal healthcare information, which is finally transmitted to the edge server via 5G cellular network. To support a large number of IoMT devices simultaneously, the method of Device-to-device (D2D) communication is introduced into 5G cellular network to utilize direct links like Bluetooth or WiFi-direct between neighbor devices [7]. D2D is regarded as one of the most promising technologies for future 5G network since it can improve spectrum utilization, increase network capacity, and reduce communication latency [8].

D2D has directly served the communication ability for 5G enabled IoMT. However, the advantage comes with the cost of other security risks that have never been considered in the conventional IoMT system [1]. The information that is sharing between the medical experts and the patients is so crucial and sensitive that secure transmission is much necessitated [9]. Due to the open nature of the wireless medium, 5G and D2D enabled IoMT system is vulnerable to various attacks, such as man-in-the-middle attack, replay attack, denial-of-service attack, and impersonation attack [10]. Traditional cryptography and authentication based secure mechanism can protect the IoMT system against the attacks from external abnormal and illegitimate entities. However, the adversary can disable the authentication system by capturing and manipulating the legitimate IoMT devices. Hence, despite of identity protection via authentication system, internal compromised devices cannot be isolated from the system through traditional security mechanisms [11]. Trust management system is one significant security solution against the aforementioned issues [12]. The general definition of trust is "confidence in or reliance on some quality or attribute of a person or thing, or the truth in a statement" [13]. The introduction of trust mechanism in IoMT system can help to isolate the adverse or selfish devices by trust

Manuscript received July 31, 2021; revised September 30 and November 2, 2021; accepted November 8, 2021. This work was supported in part by the National Natural Science Foundation of China under Grant 61801072, in part by the Science and Technology Research Program of Chongqing Municipal Education Commission under Grant KJQN202000641, in part by the Natural Science Foundation of Chongqing under Grant cstc2020jcyj-msxmX0636, and in part by the Japan Society for the Promotion of Science (JSPS) Grants-in-Aid for Scientific Research (KAKENHI) under Grants JP18K18044 and JP21K17736. (*Corresponding authors*: *Keping Yu; Simon X. Yang.*)

Liu Yang, Yinzhi Lu, and Tan Guo are with the School of Communication and Information Engineering, Chongqing University of Posts and Telecommunications, Chongqing, China (e-mail: yangliu@cqupt.edu.cn; henanluyinzhi@163.com; guot@cqupt.edu.cn).

Keping Yu is with Global Information and Telecommunication Institute, Waseda University, Shinjuku, Tokyo 169-8050, Japan (e-mail: keping.yu@aoni.waseda.jp).

Simon X. Yang is with the Advanced Robotics and Intelligent Systems Laboratory, School of Engineering, University of Guelph, Guelph, ON N1G2W1, Canada (e-mail: syang@uoguelph.ca).

Chinmay Chakraborty is with Birla Institute of Technology, Mesra, India (email: cchakrabarty@bitmesra.ac.in).

evaluating and then, improve the cooperation between devices and enhance the security performance of the system [14], [15].

### A. Motivation

5G makes the real-time healthcare monitoring possible while D2D communication can greatly offload the mobile traffic from the base station. However, direct communication between IoMT devices is vulnerable to the attacks from manipulated devices. To detect the malicious IoMT devices, the reliability of each device needs to be measured through trust evaluating according to the monitored transmission behaviors [16]. Hence, IoMT devices can avoid direct communications with the malicious ones by introducing the trust management technique. However, the monitored transmissions are with some uncertainties due to accidental network fault or disturbance arising from the unreliable system and network environment [17], [18]. For instance, if a device does not overhear a forwarded packet after it transmits data to another device in proximity, a malicious dropping or missed overhearing may occur that cannot be identified. Similarly, it is hard to distinguish a malicious delaying from a normal one caused by retransmission if a delayed transmission event is captured. The transmission uncertainty can significantly affect the accuracy of trust evaluation. Furthermore, it is with great challenges to deal with the uncertainty problem in an open wireless medium, since the dynamic characteristics make the probability of missed overhearing or the rate of retransmission unstable. To solve these problems, in this paper we present an intelligent trust cloud management (ITCM) method for 5G and D2D enabled IoMT system. The characteristics of the open wireless medium can be learned dynamically while the trust relationship between IoMT devices can be timely updated.

### B. Contribution

The main contributions are shown as follows:

1) We present a novel intelligent trust cloud management method for 5G edge computing and D2D enabled IoMT system under a dynamic communication environment.

2) To address the trust uncertainty issue, fuzzy logic theory is adopted to estimate the trust value of an IoMT device. Moreover, cloud model is further used to cluster multiple trust values of a device to make the trust evaluation more reliable.

3) To make the trust management system intelligent and adaptive, a cooperative labeling and training mechanism is adopted to construct the standard trust cloud frameworks.

4) A trust classification method is presented to determine whether an IoMT device is malicious or not. The device can be classified into either the malicious group or the normal one according to the individual trust cloud of this device and the trained standard trust clouds.

5) A trust cloud update scheme is presented which makes the proposed trust management method adapt to the dynamic characteristics of an open wireless medium.

6) For secure and efficient data transmission from IoMT device to edge server, the proposed trust management method is applied to clustering while its performance is evaluated via simulation experiments.

## II. RELATED WORK

Recently, trust management has been widely adopted to address the security issue in IoT field. According to the trust evaluation methods, trust management systems can be divided into the traditional system and the one with intelligence.

### A. Traditional trust management systems

To avoid black hole attack in routing process, Liu *et al.* [19] proposed an active detection-based security and trust routing method called ActiveTrust. To improve the data transmission security, numbers of detection routes were created to quickly acquire individual trust value. Moreover, the generation and distribution of detection routes were given in an active mode that can enhance the energy efficiency. ActiveTrust can rapidly detect the malicious attackers while assuring transmission security. However, extra cost was necessary to actively generate the detection routes.

A Beta-based trust management (BTM) mechanism was presented to defend internal attacks for sensor networks [20]. Firstly, direct trust value was calculated according to Beta distribution where communication factors were used as the shape parameters. Secondly, the final trust value was computed by weighing the direct trust value and recommended one. Finally, based on the trust variations, threshold, and window length, a malicious detection method triggered by time window was proposed to isolate betrayal nodes from the network. To some extent, BTM can defend against bad mouthing attacks since the numbers of right and wrong trust estimations were used to calculate the recommended trust value. But experiment results showed that a quick decline of trust level did not occur for continuous malicious behaviors that may make the trust management system vulnerable.

To solve the trust uncertainty problem in an open wireless medium, Zhang *et al.* [21] presented a trust evaluation method for clustered Wireless Sensor Networks (WSNs) based on cloud model (TECC). Firstly, multiple factors that include communication, message, and energy were used to establish the corresponding factor trust clouds. Secondly, immediate trust cloud was computed by weighing these factor trust clouds. Thirdly, the final trust cloud was constructed by synthesizing the immediate trust cloud and the recommended one according to a time sensitive factor. Finally, the final trust cloud was converted into grade level through cloud decision-making. TECC can effectively address the trust uncertainty issue due to unique characteristics of cloud model. However, the standard trust clouds were previously designed based on trust experience which may not always be effective in a dynamic environment.

A novel trust cloud model (TCM) was presented to evaluate the trust of sensor nodes in underwater WSNs [22]. The procedure of TCM was composed of communication behaviors monitoring, direct trust cloud constructing, weighing of multiple trust clouds, and similarity judgement between the weighted trust cloud and standard ones. Owing to the cloud model, TCM can effective solve the trust uncertainty problem. However, how to establish the standard trust clouds was not referred in TCM.

TABLE I
COMPARATIVE ANALYSIS

| Trust model | Model construction method | Outlier detection method | Trust uncertainty consideration | Adaptability to dynamic environment |
|---|---|---|---|---|
| ActiveTrust [19] | Weighing | Trust boundary | N | Strong |
| BTM [20] | Beta distribution, weighing | Trust boundary | N | Strong |
| TECC [21] | Cloud model, weighing | Trust boundary | Y | Weak |
| TCM [22] | Cloud model, weighing | Trust classification | Y | Weak |
| GDTMS [23] | Gaussian distribution | Grey theory | N | Strong |
| CPMAED [24] | K-Means, SVM | Trust classification | N | Weak |
| NMLTM [25] | Weighing, K-NN, SVM | Trust boundary | Y | Medium |
| MLTC [14] | PCA, K-Means, SVM | Trust boundary | N | Medium |
| STMS [26] | K-Means, SVM | Trust boundary | N | Medium |
| TEUC [27] | C4.5 decision tree | Trust classification | Y | Medium |
| Proposed ITCM | Cloud model, adaptive learning | Trust classification | Y | Strong |

Fang et al. [23] proposed a Gaussian distribution-based comprehensive trust management system (GDTMS) for fog computing based WSNs. To update the trust relationship between sensor nodes, the Gaussian distribution based trust model was constructed based on historical interactions. And then a grey decision-making scheme was introduced to detect the reliable and energy-efficient routes. GDTMS can achieve a trade-off among security, transmission performance, and energy. However, simulation results showed that the trust level was gradually decreasing for continuous noncooperation that may make the trust management system vulnerable.

*B. Intelligent trust management systems*

Liu et al. [24] presented an efficient detection framework against conditional packet manipulation attack (CPMAED) for IoT networks. Firstly, a regression model was trained to estimate the trust values of sensor nodes according to the reputation of routing paths. Secondly, the estimated trust values were clustered into three groups with different trust grades. Finally, whether a node was benign or malicious can be determined according to the group to which its trust value belongs. Simulation results verified that CPMAED can effectively prevent against conditional packet manipulation attack. However, the trust value of a node can be influenced by the behaviors of other nodes along the same routing path, which may reduce the detection accuracy of malicious nodes.

Khan et al. [25] proposed a neutrosophic machine learning based trust model (NMLTM) for industrial Internet of Things (IIoT) applications. A neutrosophic weighted product method was presented to compute the trust scores of IIoT devices according to the characteristics that include spatial knowledge, temporal experience, and behavior pattern. Based on the trust scores, neutrosophic K-NN clustering method was adopted to label the characteristics of IIoT devices. Neutrosophic SVM algorithm was finally used to determine the best trust threshold so that whether an IIoT device was trustworthy or not can be identified. The proposed trust model can effectively decrease the impact of outliers since the trust scores were further analyzed via machine learning technique. However, the dynamic nature of industrial field was not considered that may degrade the performance of the trust model.

A machine learning based trust computational (MLTC) model was presented to acquire secure IoT services [14]. Firstly, trust attributes from the aspects of knowledge, experience, and reputation were calculated while the features of these attributes were extracted. And then, PCA algorithm was adopted to reduce the dimension of the features that were further used to label the trust attributes according to K-Means algorithm. Finally, the labeled trust attributes were used to train a SVM based trust prediction model that can identify the trust boundaries for future interactions. MLTC had skillfully addressed the trust prediction issue using machine learning method. However, the trained SVM framework was not further updated during the operation of the system that may affect the security performance.

A synergetic trust model using SVM (STMS) was proposed for underwater WSNs [26]. Sensor nodes in the network were grouped into clusters. Some member nodes and two cluster heads were included in each cluster. One cluster head acted as the master (MCH), and the other was the slave (SCH). Trust attributes regarding communication, packet, and energy were periodically collected by each cluster member and finally transmitted to MCH. K-means was adopted by MCH to mark these trust attributes with labels good (1) or bad (0). Moreover, a SVM trust classification framework was further trained. The result was sent back to cluster members for malicious nodes detection. To determine whether a MCH was malicious, its trust value was calculated by the SCH based on the trust attributes. Once a MCH was considered as the malicious, it would be replaced. Supervised and unsupervised learning methods were successfully adopted to construct the trust management system in STMS. Nevertheless, the selection of double cluster heads required extra energy overhead.

A dynamic trust evaluation and update method based on C4.5 decision tree (TEUC) was presented for underwater WSNs [27]. The trust evidences were collected firstly including transmission delay, successful interactions, residual energy, and packet consistency. Then these evidences were normalized and discretized to be fuzzy sets. The fuzzy trust evidences were further used to train a C4.5 decision tree for trust evaluation. Finally, the reward and penalty factors were defined to update the trust based on a sliding time window. Machine learning algorithm had been successfully adopted for trust evaluation in TEUC, and a good performance of malicious nodes detection had been achieved according to the simulation results. C4.5 is a kind of supervised learning algorithm, the collected trust evidences at the beginning of network operation was used to train the decision tree since it was assumed that no malicious attacks occur in this stage. However, dynamic underwater environment may make the initially trained decision tree unsuitable in the future network operation phase.

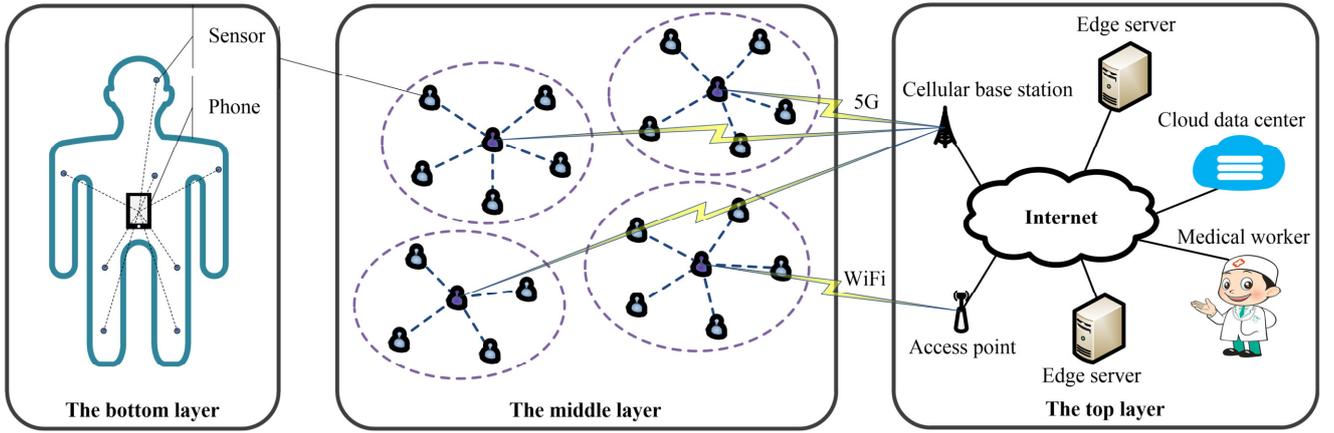

Fig. 1. 5G edge computing and D2D enabled IoMT framework.

To show the difference between our proposed method ITCM and the related trust models, we give the comparative analysis results that can be seen in Table I.

## III. SYSTEM MODEL

We use the same radio model and energy model as those in our previous work [17]. In the rest of this section, the IoMT framework, cloud model, trust management model, and some assumptions are described in detail.

### A. IoMT framework

Three layers are included in the network model of the 5G edge computing and D2D enabled IoMT framework, as shown in Fig. 1. The bottom layer is a typical WBAN that consists of several sensors and a smart phone. The healthcare information of a patient is collected by the sensors and then directly transmitted to the phone due to the short distance. In the middle layer, large numbers of WBAN entities are grouped into clusters to cooperatively transmit the information. Each cluster contains a head and some members. A member entity transmits information to its own head via D2D communication while a head transmits the information to base station or access point through 5G or WiFi communications. Finally, the healthcare information is transmitted to the nearby edge server in the top layer. The medical results are delivered to the cloud data center if necessary.

### B. Cloud model

The mathematical concept cloud is a cognitive model that describes the uncertainty of linguistic concepts. Especially, it can deal with the fuzziness and randomness by transforming linguistic concepts into quantitative values [21]. A typical cloud consists of large numbers of cloud drops. A single drop is meaningless, whereas a cloud with a lot of drops expresses the feature of a qualitative concept. A cloud model describes the transformation according to the following three numerical characteristics [22]:

1) Expectation $Ex$ indicates the mathematical expectation of the cloud drops that belonging to a qualitative concept in the universal.

2) Entropy $En$ measures the uncertainty of the concept.

3) Hyper entropy $He$ represents the uncertainty grade of the entropy $En$.

Normal cloud model is a typical one which is defined based on normal distribution. Let $d$ be a random drop instance of a normal cloud with the numerical characteristics $Ex$, $En$, and $He$, then the following conditions are satisfied: 1) $d \sim N(Ex, En'^2)$; 2) $En' \sim N(En, He^2)$.

### C. Trust cloud management model

To address the trust uncertainty issue and adaptively update the trust relationship between IoMT devices under a dynamic network circumstance, we propose an intelligent trust cloud management method named ITCM. As shown in Fig. 2, the model of ITCM consists of standard trust clouds training, individual trust clouds constructing, trust decision-making, and trust clouds updating modules. To explore the dynamic characteristics of an open wireless medium, IoMT devices are cooperatively marked as the malicious or normal ones in the initial stage of the system deployment. Standard trust clouds regarding the normal and malicious populations can be established respectively according to the estimated trust values of the marked IoMT devices. With the system operating, the trust cloud of any individual IoMT device, called the individual trust cloud for short, is constructed via fuzzy trust inferring and trust recommending. Through trust decision-making, an IoMT device is recognized to be either malicious or normal according to the trust classification result. With the interacting between IoMT devices, the standard and individual trust clouds are updated to adapt to the dynamic characteristics of an open wireless medium.

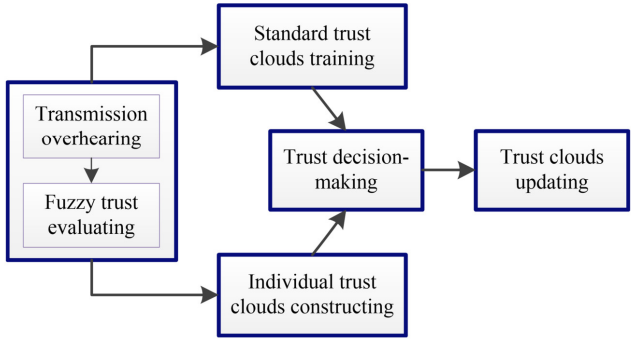

Fig. 2. Our intelligent trust cloud management model.

*D. Assumptions*

Before we describe the detail of our trust management method, several assumptions regarding the communication medium and IoMT devices are given as follows:

1) The quality of the wireless communication medium is not stable due to accidental default or unpredictable interference from external environment.

2) The captured or manipulated IoMT devices become the compromised devices with heterogeneous attack capabilities.

3) The energy of any IoMT device to be used for healthcare monitoring is limited since most of the energy has to be used for other aspects.

## IV. THE PROPOSED INTELLIGENT TRUST CLOUD MANAGEMENT METHOD

Usually, a fuzzy logic system can be used to address the trust uncertainty issue when determining the trust value of an IoMT device. However, a single trust value cannot always indicate the reliability of a device with high precision. Then a cloud model is introduced to cluster multiple trust values of a device that can improve the accuracy of trust estimation. To make the trust management intelligent and adaptive, a cooperative labeling and training mechanism is adopted to construct the standard trust cloud frameworks. By performing trust classification, whether an IoMT device is malicious or not can be determined. The detail of the proposed trust management method is given in the rest of this section.

*A. Trust evidences collecting and inferring*

To estimate the trust value of an IoMT device, the trust evidences regarding the behaviors of this device need to be collected firstly. Since our purpose is assuring the secure and reliable data transmission between IoMT devices, the trust evidences that reflect the changes of data-plane information when attacks happen should be collected. Usually, tempering, dropping, and delaying events are considered as the trust evidences, which can be gathered through transmission overhearing. Due to the failure of packet authentication, a tampered packet will be dropped directly so that a tampering event can be somehow regarded as a dropping event.

Once the trust evidences have been collected, trust attributes including the timely forwarding rate (TFR) and successfully forwarding rate (SFR) can be computed. Then the trust values of the related IoMT devices can be estimated using an interval type-2 fuzzy logic system [17]. For any input vector which consists of the current trust attributes of an IoMT device, the trust value of this device can be computed through fuzzy inferring according to the rules mapping the fuzzy sets between input trust attributes and output trust value. Firstly, the interval type-2 fuzzy sets corresponding to the trust attributes are acquired for each fuzzy rule. Secondly, some interval type-2 fuzzy sets corresponding to the output trust value are inferred according to the rules. By performing type reduction, a type-1 fuzzy set corresponding to the trust value is obtained. Finally, the output trust value is estimated through defuzzification.

*B. Standard trust clouds training*

Generally, no malicious or compromised IoMT devices exist during the initial stage of the system deployment. Each IoMT device explores the characteristics of the open wireless medium including packet dropping and delaying rates at this stage. Through active transactions with neighbors, the standard trust cloud frameworks can be trained. The training process is individually initiated by each IoMT device and then operates based on rounds.

For any IoMT device $i$, once it starts the training process of its own standard trust clouds, it randomly selects a router $j$ and destination $k$ within the corresponding neighborhood radius $R_n$ during each training round. The router $j$ firstly accepts a malicious label and then a normal one to forward data packets for the number $N_f$ times separately. If $j$ is labeled as the malicious, it drops packets with the probability $P_{dp}$ and gives a delay with the probability $P_{dy}$. These attack probabilities should be obviously smaller than the packet loss or retransmission probabilities to highlight the trust uncertainty. The malicious delay is a random one with the maximum duration $MAX\_DUR$. Once $i$ transmits a data packet to $j$, it keeps overhearing the followed forwarding until the packet is overheard or the maximum duration is up. Due to the dynamic characteristics of an open wireless medium, packet loss or missed overhearing may occur. If $j$ is labeled as a normal IoMT device and does not get a reply from destination $k$ after forwarded the packet, it retransmits the packet after a delay. A dropping event is recorded by $i$ if it does not overhear the forwarded packet within a specific time slot. In addition, a delaying event is recorded if a malicious delaying or normal retransmission event is captured. Based on the overheard transmission behaviors, device $i$ evaluates the trust value of $j$ via the interval type-2 fuzzy logic system.

After each round of active training, the initiator device $i$ can acquire the number $N_f$ of trust values for both the labeled malicious devices and normal ones. These trust values form a labelled training dataset that can be used to establish the malicious and normal standard trust cloud frameworks. To construct the standard trust clouds, the data in the training dataset can be regarded as cloud drops. If the numbers of both the malicious drops and the normal ones reach the maximum number $MAX\_DRP$ after one round of training, the two kinds of drops can be used to establish the initial malicious and normal standard trust clouds respectively. Let $STC_m(i) = stc_i(Ex_m, En_m, He_m)$ denote the malicious standard trust cloud constructed by $i$, and $STC_n(i) = stc_i(Ex_n, En_n, He_n)$ be the normal one. Let $D$ denote the trust cloud drop, and $n$ be the total number of drops, then the three numerical characteristics of the trust cloud can be calculated as follows [22]:

$$Ex = \frac{1}{n}\sum_{l=1}^{n} D_l. \qquad (1)$$

$$En = \sqrt{\frac{\pi}{2}} \frac{1}{n}\sum_{l=1}^{n} |D_l - Ex|. \qquad (2)$$

$$He = \sqrt{\left|\frac{1}{n-1}\sum_{l=1}^{n}(D_l - Ex)^2 - En^2\right|}. \qquad (3)$$

After the initial standard trust clouds have been established, the training process continues until an obvious classification boundary between the two standard trust clouds occurs. Then the condition $Ex_m < Ex_n$ should be satisfied. Otherwise, the training process is forced to terminate when the total number of training rounds reaches the maximum $MAX\_TR$. If an IoMT device finishes its active training, it recommends the trained standard trust clouds to its neighbors. After an IoMT device has received all the recommendations from neighbors, it takes the average as its own final standard trust clouds, which are denoted by $STC'_m$ and $STC'_n$ for the malicious and normal ones respectively.

### C. Individual trust clouds constructing

To determine whether an IoMT device can be trusted or not, the individual trust cloud of this device needs to be constructed firstly. In the clustered IoMT system, a normal IoMT device transmits the data to its own cluster head device and then overhears the followed transmission behaviors of the head. Once the trust evidences have been collected, the trust value of this cluster head device can be estimated through fuzzy inferring and then considered as the cloud drop for individual trust cloud construction. However, an IoMT device estimates the trust value of another device only when the two devices are grouped into the same cluster while one is a member and another is the head. To efficiently acquire trust information of the IoMT devices, the mechanism of trust recommendation is introduced into our trust management method. If an IoMT device has chosen a trusted device as its own head, then it can send the trust recommendation request to its head. To save the energy, a device does not send the request to other devices.

Let $\overline{T}_{i,j}$ denote the average trust value of the device $i$ to $j$, and $\overline{T}_{j,k}$ represent the recommended average trust value from $j$ to $k$, then $i$ can update the trust value $T_{i,k}$ to $k$ as follows:

$$T_{i,k} = \begin{cases} \left(\overline{T}_{i,k} + \overline{T}_{j,k}\overline{T}_{i,j}\right)/\left(1+\overline{T}_{i,j}\right) & \text{if } \overline{T}_{i,k} > 0 \\ \overline{T}_{j,k}\overline{T}_{i,j} & \text{otherwise} \end{cases}, \qquad (4)$$

where $\overline{T}_{i,k}$ is the average trust value of $i$ to $k$.

For any IoMT device $i$, once it updates the trust to $j$ through fuzzy inferring or recommending, it stores it into a trust cloud drop set $S_{TCD}(i,j)$. If the set size reaches the threshold $THR\_DRP$, the individual trust cloud $ITC(i,j)$ of $i$ to $j$ can be constructed according to the drops and the equations 1-3.

### D. Trust decision-making

To construct the secure clusters where malicious IoMT devices are isolated from being the heads, a normal IoMT device usually needs to select a trusted head to join in cluster. For any IoMT device $i$, it can identify whether a cluster head device $j$ is malicious or not by performing trust classification according to the corresponding individual trust cloud and the final standard ones. If the expectation of the individual trust cloud $ITC(i,j)$ is obviously smaller than that of the final standard one $STC'_m(i)$, then the cluster head device $j$ is classified into the malicious group. If the expectation of $ITC(i,j)$ is obviously bigger than that of $STC'_n(i)$, then the cluster head device $j$ can be classified into the normal group. Otherwise, the similarities between the individual trust cloud and the final standard ones have to be computed to determine the group to which the device $j$ belongs. If the cloud $ITC(i,j)$ is more similar to the one $STC'_m(i)$, then the cluster head device $j$ is considered as the malicious head.

To estimate the similarity between an individual trust cloud and a final standard one, a certain number $N_{DRP}$ of drops are randomly generated based on the individual trust cloud. The average degree of these drops belonging to the standard trust cloud is regarded as the similarity between the two trust clouds. To generate a random individual trust cloud drop and compute its membership degree, four steps are performed as follows:

1) A random standard deviation $\sigma_n$ of the individual trust cloud is generated according to a normal distribution, where the mean is the fuzziness entropy of the individual trust cloud, and the standard deviation is the uncertain degree of this fuzziness entropy.

2) The individual trust cloud drop $D_I$ can be generated according to a normal distribution, where the mean is the expectation of the individual trust cloud, and the standard deviation is $\sigma_n$.

3) A random standard deviation $\sigma_s$ of the standard trust cloud is generated according to a normal distribution, where the fuzziness entropy of the standard trust cloud acts as the mean while the uncertainty of the fuzziness entropy is the standard deviation.

4) The membership degree $M_{ITD}$ of the individual trust cloud drop $D_I$ can be calculated by

$$M_{ITD} = \exp\left(-(D_I - Ex_s)^2 / (2\sigma_s^2)\right), \qquad (5)$$

where $Ex_s$ is the expectation of the standard trust cloud.

*E. Trust clouds updating*

With the system operating, the quality of wireless medium may change due to the aggravation or mitigation of interference from external environment. Then the initially trained standard trust clouds may not always be applicable. The dynamic characteristics of the wireless medium may also cause the invalidation of the individual trust clouds. Hence, both the standard trust clouds and individual ones need to be updated in time to adapt to an open wireless medium.

Once an IoMT device $i$ has updated the trust values of its neighbors, it reconstructs the individual trust cloud for each neighbor $j$ based on the drops in the set $S_{TCD}(i,j)$, which preserves the number $THR\_DRP$ of the latest trust values. According to the reconstructed individual trust clouds of neighbors and the final standard ones, whether these neighbors are malicious or not can be determined. The updated trust values of the neighbors that are classified into the malicious group are added into a malicious cloud drop set. Meanwhile, the updated trust values of other neighbors are added into a normal cloud drop set. If the size of either cloud drop set reaches the maximum $MAX\_DRP$, a new standard trust cloud $STC^*(i)$ can be established based on the cloud drops. Then the IoMT device $i$ can update its final standard trust cloud $STC'(i)$ as follows:

$$STC'(i) = \alpha STC'(i) + \beta STC^*(i), \qquad (6)$$

where $\alpha$ and $\beta$ are the weighing factors corresponding to the previous final standard trust cloud and the current standard trust cloud respectively.

Since the current standard trust cloud is trained by using a dataset which is constructed in a short period of time, it should be given a much smaller weight than the previous final standard trust cloud to make the update smooth. Once an update is completed, all drops in the corresponding malicious or normal cloud drop set are cleared.

## V. THE DETAIL OF OUR SECURE CLUSTERING PROTOCOL

The procedure of our secure clustering method is given in Fig. 3. Initially, IoMT devices individually initiate the standard trust clouds training process. After that, the secure clustering process works based on rounds, including the courses of secure clusters formation, data transmission, transmission overhearing, individual trust clouds constructing, trust decision-making, and standard trust clouds updating.

To construct the secure clusters, IoMT devices firstly self-decide whether to be the heads. After the cluster heads have been selected, each normal IoMT device selects a trusted one to join in cluster. If an IoMT device has the eligibility to be the head in the current round $r$, it self-decides whether to be a head based on the threshold $Thr$ which is computed as follows:

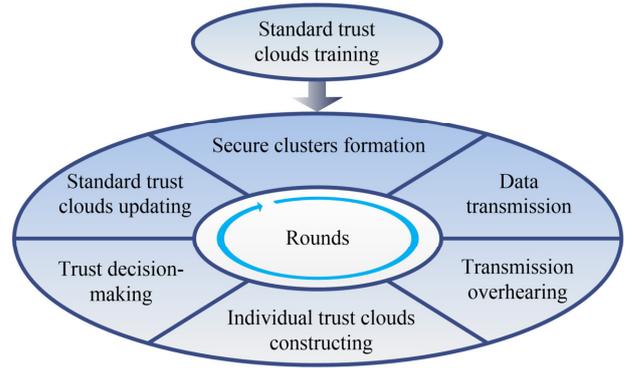

Fig. 3. The procedure of the proposed secure clustering protocol.

$$Thr = p_{CH} / (1 - p_{CH}(r \bmod (1/p_{CH}))), \qquad (7)$$

where $p_{CH}$ is the probability of being the head. If an IoMT device has not been the head during the latest $1/p_{CH}$ rounds, then it has the eligibility of being a head in the current round.

If an IoMT device decides to be a head, it informs the election to its neighbors. To construct the secure clusters, IoMT devices should avoid joining in the clusters where the corresponding heads are detected to be malicious. If a normal IoMT device receives a head election message, it adds the message broadcaster as its own head candidate. After trust decision-making, it chooses the nearest trusted candidate to join in the cluster. If all individual trust clouds with regard to the candidates are not constructed due to insufficient trust cloud drops, then the nearer candidate which has not been interacted with or the one that is estimated to be the most reliable is selected. If a normal IoMT device finally does not find an appropriate cluster to join in, then it has to decide to be a head if it is eligible in the current round.

After all clusters are grouped, each member IoMT device begins to collect healthcare information of the patient and then transmits it to the head during a specially allocated time slot. By overhearing the transmission behaviors, each member IoMT device calculates the trust value of the head according to interval type-2 fuzzy logic. Through fuzzy trust inferring or trust recommending, an IoMT device updates the trust values of others while constructing or reconstructing the individual trust clouds if the related cloud drops are sufficient. After trust decision-making, an IoMT device is malicious or not can be detected. The latest trust value of this device is used to update the standard trust clouds.

## VI. PERFORMANCE EVALUATION

*A. Simulation setup*

To model the open wireless medium, a Markov chain is adopted to describe the channel quality which may be bad or good with the probabilities $p_0$ and $1-p_0$ respectively [17]. Here $p_0$ can be calculated as follows:

$$p_0 = \frac{\alpha_0}{\alpha_0 + \alpha_1}, \qquad (8)$$

TABLE II
PARAMETERS SETTING

| Parameter | value | Parameter | value |
| --- | --- | --- | --- |
| Packet size (bits) | 3000 | $R_n$ (m) | 25 |
| Control packet (bits) | 300 | $N_f$ | 20 |
| Training packet (bits) | 300 | $P_{dp}$ | 0.05 |
| Initial energy $E_0$ (J) | 1 | $P_{dy}$ | 0.05 |
| $E_{elec}$ (nJ/bit) | 50 | MAX_DUR (s) | 10 |
| $\varepsilon_{amp}$ (pJ/bit/m$^4$) | 0.0013 | MAX_DRP | 100 |
| $\varepsilon_{fs}$ (pJ/bit/m$^2$) | 10 | MAX_TR | 20 |
| $E_{DA}$ (nJ/bit/message) | 5 | THR_DRP | 20 |
| $E_h$ (nJ/bit) | 5 | $N_{DRP}$ | 50 |
| $E_m$ (nJ/s) | 10 | $\alpha$ | 0.8 |
| $p_{CH}$ | 0.07 | $\beta$ | 0.2 |

where $\alpha_0$ and $\alpha_1$ denote the rates of bad and good states respectively.

Initially, we set $\alpha_0$ and $\alpha_1$ to be 1 and 9 respectively. To describe the dynamic characteristics of an open wireless medium, $\alpha_0$ and $\alpha_1$ are set to be 2 and 8 after some rounds of data transmission, and set to be 3 and 7 after some more rounds. Finally, $\alpha_0$ and $\alpha_1$ restore to the initial values. Since the locations of IoMT devices are usually different, we assume that the channel quality is independent for each device. If interference occurs, the channel quality becomes bad so that the data packet cannot be received or overheard successfully.

The heterogeneous malicious IoMT devices can be divided into the generic, advanced, and super groups that account for 30%, 40%, and 30% of the total number of malicious devices. The three kinds of malicious devices drop the packets during the data transmission process with the probabilities $2P_{dp}$, $4P_{dp}$, and $6P_{dp}$ respectively. In addition, they launch delaying attacks with the probabilities $2P_{dy}$, $4P_{dy}$, and $6P_{dy}$ respectively.

In our proposed method ITCM, fuzzy logic and cloud model are used to address the trust uncertainty issue. The trust framework is constructed and dynamically updated by learning the characteristics of the open wireless medium. To evaluate the performance of ITCM, we compare it with that of TECC and TEUC in secure clustering applications. This is because that cloud model is also adopted in TECC to deal with the trust uncertainty problem. Although machine learning technique is used in TEUC for trust evaluating while the fuzziness of trust evidences is still considered. However, the initially trained trust model is not updated in the followed steady operation of the system. For the development of performance comparison, TFR and SFR are used as the trust attributes in TECC and TEUC. The trust recommendation method is also added into TEUC for consistency. In addition, the failure tolerance factor is 0.2 and time sensitive factor is 0.6 in TECC. The higher and lower thresholds that are used to divide the trust evidences into three fuzzy levels are 0.9 and 0.7 in TEUC. Other parameters used in the experiments are listed in Table II.

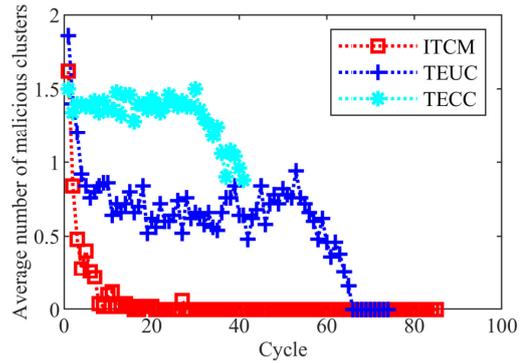
Fig. 4. Average number of malicious clusters.

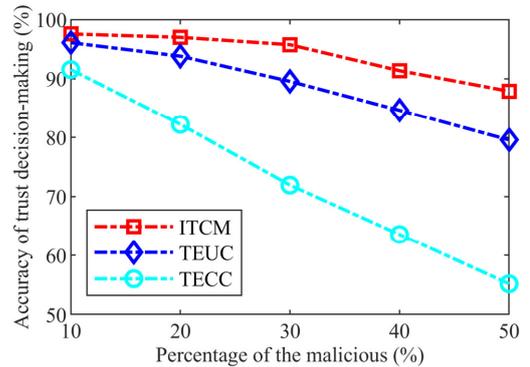
Fig. 5. Accuracy of trust decision-making with respect to different percentages of the malicious.

B. *Experiment results*

In this section, we verify the performance of the intelligent trust cloud management method ITCM via MATLAB platform. The simulation experiment is firstly performed under the case that 100 IoMT devices are randomly deployed in a rectangular sports area with the scale of 100 × 100 m$^2$. The percentage of the malicious IoMT devices ranges from 10 to 50. To evaluate the scalability of ITCM, the experiment is further conducted for the IoMT systems with different network sizes and numbers of devices. The percentage of malicious IoMT devices is 30. All experiments are repeated 100 times independently, and we take the average as the final results. In addition, some key metrics are selected to estimate the corresponding confidence intervals, which can be seen in the following result analysis.

The comparison on average number of malicious clusters in each cycle is shown in Fig. 4. A cycle is totally 50 data transmission rounds, and the percentage of the malicious device is 20. This figure shows that the numbers of given cycles for the three methods are not the same. TECC has the fewest while our method ITCM has the most. This is because the number of cycles directly depends on the lifetime of the last dead IoMT device due to energy depletion. This figure also shows that the number of malicious clusters in ITCM decreases much faster than that in TECC and TEUC. In addition, the number in TEUC decreases faster than that in TECC. These cases indicate that ITCM has the best performance on preventing malicious devices against being selected as cluster heads, and followed by TEUC. For the proposed method ITCM, malicious IoMT devices no longer have the chance to be the cluster heads after some cycles of data transmission. Hence, ITCM is very suitable for the industrial applications with high security requirements.

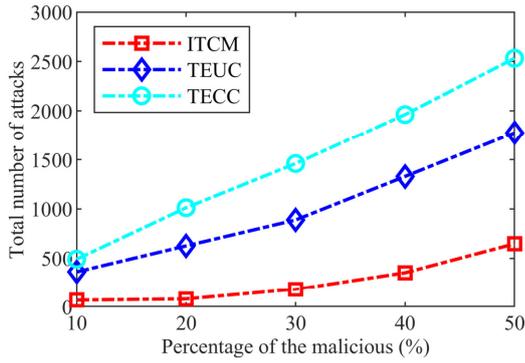

Fig. 6. Total number of attacks with respect to different percentages of the malicious.

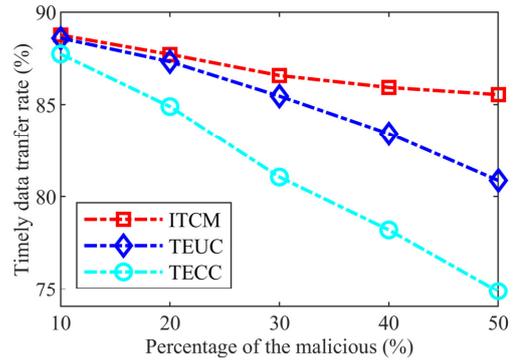

Fig. 8. Timely data transfer rate of member devices with respect to different percentages of the malicious.

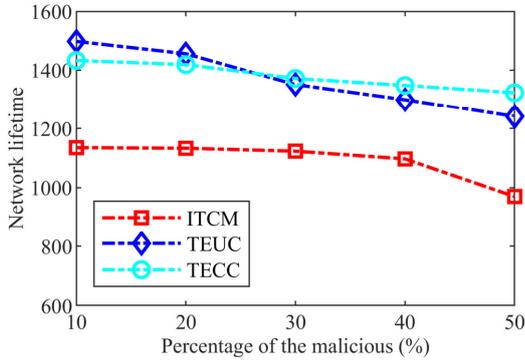

Fig. 7. Network lifetime with respect to different percentages of the malicious.

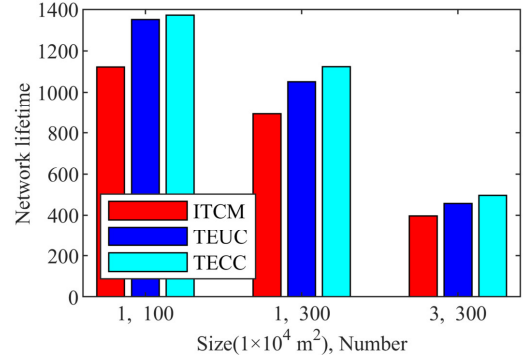

Fig. 9. Network lifetime with respect to different network sizes and numbers of IoMT devices.

The comparison on trust decision-making accuracy among TECC, TEUC, and ITCM with respect to different percentages of the malicious is given in Fig. 5. It shows that our method ITCM has the highest accuracy of trust decision-making while TECC has the lowest for all the percentage instances of malicious devices. This is because fuzzy logic system and cloud model are used to address the trust uncertainty issue in ITCM. Moreover, the dynamic characteristics of an open wireless medium are explored to adaptively train the standard trust clouds. Hence, the trust-decision making accuracy of the proposed method is 87.9% (the 95% confidence interval: 83.1 - 92.7) even when the percentage of the malicious is 50. Cloud model is also adopted in TECC to construct the trust clouds. However, the standard trust clouds are established in advance with experience. Although the trust decision tree is trained by using machine learning algorithm in TEUC, it is not timely updated to adapt to the dynamic wireless medium. This figure also shows that with the increase of malicious device percentage, the trust-decision making accuracy decreases due to more malicious devices in the IoMT system.

The comparison on total number of attacks among TECC, TEUC, and ITCM is given in Fig. 6. From this figure, we can see that when the percentage of the malicious increases, the total numbers of attacks increase simultaneously for these methods. Since our method ITCM has the highest accuracy of trust decision-making, it has the fewest attacks for all the percentage instances of malicious devices, as shown in this figure. The total number of attacks of the proposed method is 647 (the 95% confidence interval: 561 - 733) even when the percentage of the malicious is 50. Because TEUC has higher decision accuracy than TECC, it has fewer attacks that can also be seen in this figure.

The comparison on network lifetime among TECC, TEUC, and ITCM is given in Fig. 7. We define the network lifetime as total alive rounds of the first normal IoMT device that exhausts the energy. This figure shows that the network lifetime decreases for the three methods when the percentage of the malicious devices increases. This is because more malicious devices are isolated from being the heads when the percentage increases, then the normal ones have more chances to be selected as the heads and then provide data forwarding services for others. ITCM has shorter network lifetime than TECC and TEUC since it has higher decision accuracy and isolates much more malicious devices from being the heads. This figure also shows that TEUC firstly has longer and then shorter network lifetime than TECC. Because that on the one hand, each IoMT device selects its own head within the entire network in TECC that may cause longer average transmission distance. On the other hand, more malicious devices are prevented against being cluster heads in TEUC that increases the forwarding burden of the normal ones.

The comparison on timely data transfer rate of member devices among TECC, TEUC, and ITCM is given in Fig. 8. Here timely data transfer rate is defined as the ratio of the total number of immediately forwarded packets to the total number of packets that need to be forwarded. This figure shows that when the percentage of the malicious increases, timely data transfer rate decreases for all these methods. Moreover, the timely data transfer rate is below the percentage 90 even when the percentage of the malicious is only 10. This is because some data dropping and retransmission events occur inevitably due to the dynamic wireless communication environment. This figure also shows that ITCM has the best result on timely data transfer rate while TECC has the worst for all the percentage instances of malicious devices. This is because malicious IoMT devices are detected with the highest accuracy and therefore have the fewest chances to launch attacks in ITCM. Whereas, malicious devices have the most chances to attack the IoMT system in TECC. The timely data transfer rate of the proposed method ITCM is 85.5% (the 95% confidence interval: 81.3 - 89.7) even when the percentage of the malicious is 50. Hence, ITCM is suitable for industrial applications with real-time requirements.

The comparison on network lifetime for different network sizes and numbers of IoMT devices is given in Fig. 9. It shows that for the networks with the same size, the one with more IoMT devices has shorter lifetime. This is because a normal IoMT device in the network with more devices has more neighbors, then it has more chances to be selected as the trusted cluster head for data transmission. This figure also shows that for the networks having the same number of IoMT devices, the one with bigger size has shorter lifetime due to the longer average communication distance.

The comparison on trust decision-making accuracy with respect to different network sizes and numbers of IoMT devices is given in Fig. 10. For all the instances of the network size and number of devices, the trust decision-making accuracy of the method TECC is obviously lower than that of TEUC and ITCM that can be seen in this figure. This case verifies that machine learning technique can effectively improve the reliability of trust evaluation. Our method ITCM has higher decision accuracy than TEUC that can also be seen in this figure. It indicates that the initially trained trust model should be updated adaptively during the steady operation phase of the system.

The comparison on total attack number for different network sizes and numbers of IoMT devices is given in Fig. 11. From this figure, we can see that our method ITCM has the fewest attacks for all the cases of network size and number of devices. This is because ITCM has the highest accuracy of trust decision-making among the three methods. For the networks with the same size, the results show that the one with more IoMT devices suffers more attacks since there are more malicious devices in the network. For the networks having the same number of devices, the results also show that the one with smaller size has more attacks for the method TECC due to the longer network lifetime. However, the converse is true for the methods TEUC and ITCM. This is because on the one hand, TEUC and ITCM have much higher accuracy of trust decision-making than TECC. On the other hand, an IoMT device in the network with smaller size has more neighbors so that it can select the trusted neighbor cluster head for data transmission with a bigger probability.

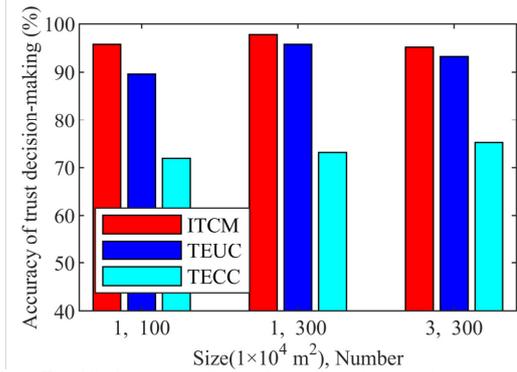

Fig. 10. Accuracy of trust decision-making with respect to different network sizes and numbers of IoMT devices.

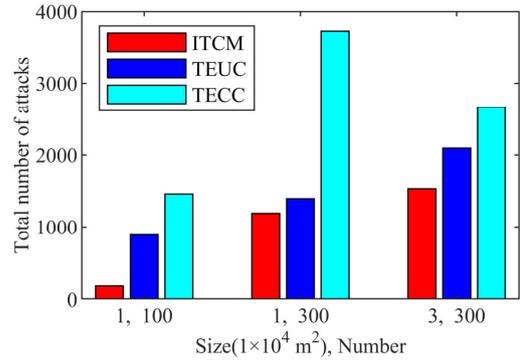

Fig. 11. Total number of attacks with respect to different network sizes and numbers of IoMT devices.

## VII. CONCLUSION

To establish the trust relationship between IoMT devices under a dynamic network circumstance, in this paper we present an intelligent trust cloud management method for 5G edge computing and D2D enabled IoMT systems. Initially, an active training process is individually performed by each IoMT device to establish the standard trust cloud frameworks. After that, the proposed method operates based on rounds including individual trust clouds constructing, trust decision-making, and trust clouds updating. The proposed method is fully distributed and easy to be implemented. Especially, it can well address the trust uncertainty issue and adapt to the dynamic characteristics of an open wireless medium. Simulation results verify that the proposed method can effectively improve the accuracy of malicious devices detection and protect the IoMT system from internal attacks.

In future works, we will further investigate the application of machine learning mechanism in trust evaluation, prediction, and management systems. Generative adversarial learning technique will be our research focus since it can generate realistic samples and distinguish real samples from fake ones. Moreover, the lightweight trust management mechanism with fast convergence speed will be investigated to satisfy the resource-constrained system.

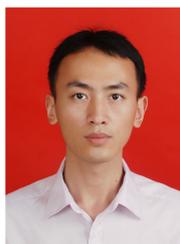

**Liu Yang** received his B.S. degree in Electronic Information Science and Technology from Qingdao University of Technology, Shandong, China, in 2010, and Ph.D. degree in Communication and Information Systems at the School of Communication Engineering, Chongqing University, Chongqing, China, in 2016. He is now a lecturer in Chongqing University of Posts and Telecommunications. His research interests include Internet of Things, data analysis, and artificial intelligence.

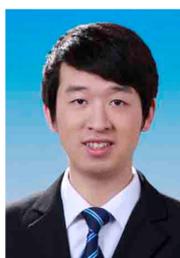

**Keping Yu** (M'17) received the M.E. and Ph.D. degrees from the Graduate School of Global Information and Telecommunication Studies, Waseda University, Tokyo, Japan, in 2012 and 2016, respectively. He was a Research Associate and a Junior Researcher with the Global Information and Telecommunication Institute, Waseda University, from 2015 to 2019 and 2019 to 2020, respectively, where he is currently a Researcher.

Dr. Yu has hosted and participated in more than ten projects, is involved in many standardization activities organized by ITU-T and ICNRG of IRTF, and has contributed to ITU-T Standards Y.3071 and Supplement 35. He is an Associate Editor of IEEE Open Journal of Vehicular Technology, Journal of Intelligent Manufacturing, Journal of Circuits, Systems and Computers. He has been a Lead Guest Editor for Sensors, Peer-to-Peer Networking and Applications, Energies, Journal of Internet Technology, Journal of Database Management, Cluster Computing, Journal of Electronic Imaging, Control Engineering Practice, Sustainable Energy Technologies and Assessments and Guest Editor for IEICE Transactions on Information and Systems, Computer Communications, IET Intelligent Transport Systems, Wireless Communications and Mobile Computing, Soft Computing, IET Systems Biology. His research interests include smart grids, information-centric networking, the Internet of Things, artificial intelligence, blockchain, and information security.

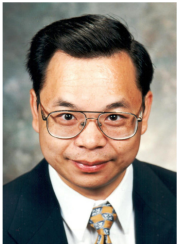

**Simon X. Yang** (S'97–M'99–SM'08) received the B.Sc. degree in engineering physics from Beijing University, Beijing, China, in 1987, the first of two M.Sc. degrees in biophysics from the Chinese Academy of Sciences, Beijing, China, in 1990, the second M.Sc. degree in electrical engineering from the University of Houston, Houston, TX, in 1996, and the Ph.D. degree in electrical and computer engineering from the University of Alberta, Edmonton, AB, Canada, in 1999.

Dr. Yang is currently a Professor and the Head of the Advanced Robotics and Intelligent Systems Laboratory at the University of Guelph, Guelph, ON, Canada. His research interests include robotics, intelligent systems, sensors and multi-sensor fusion, wireless sensor networks, control systems, machine learning, fuzzy systems, and computational neuroscience.

Prof. Yang has been very active in professional activities. He serves as the Editor-in-Chief of International Journal of Robotics and Automation, and an Associate Editor of IEEE Transactions on Cybernetics, IEEE Transactions on Artificial Intelligence, and several other journals. He has involved in the organization of many international conferences.

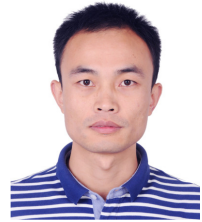

**Tan Guo** received the M.S. degree in signal and information processing and the Ph.D. degree in communication and information systems both from Chongqing University (CQU), Chongqing, China, in 2014 and 2017, respectively. Since 2018, he has been with the School of Communication and Information Engineering, Chongqing University of Posts and Telecommunications (CQUPT), Chongqing, China. He is currently a PostDoctoral Fellow with The Macau University of Science and Technology, Taipa, Macao, China. He is recipient of the Macao Young Scholars Program and the Outstanding Chinese and Foreign Youth Exchange Program of China Association of Science and Technology (CAST). His current research interests include computer vision, pattern recognition, and machine learning.

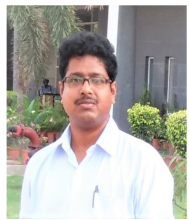

**Chinmay Chakraborty** (M'21) is an Assistant Professor in the Electronics and Communication Engineering, Birla Institute of Technology, Mesra, India and and Post-doctoral fellow of Federal University of Piauí, Brazil. He worked at the Faculty of Science and Technology, ICFAI University, Agartala, Tripura, India as a Sr. Lecturer. He worked as a Research Consultant in the Coal India project at Industrial Engineering & Management, IIT Kharagpur. He worked as a Project Coordinator of the Telecommunication Convergence Switch project under the Indo-US joint initiative. He also worked as a Network Engineer in System Administration at MISPL, India. His main research interests include the Internet of Medical Things, Wireless Body Sensor Networks, Wireless Networks, Telemedicine, m-Health/e-health, and Medical Imaging.

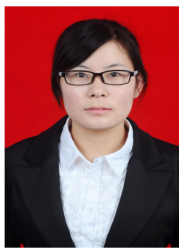

**Yinzhi Lu** received her M.S. degree in Communication and Information Systems from Chongqing University, Chongqing, China, in 2014. She is currently pursuing the Ph. D. degree in Information and Communication Engineering with the School of Communication and Information Engineering, Chongqing University of Posts and Telecommunications. She was a teaching assistant with the School of Electronic Information Engineering, Yangtze Normal University from 2014 to 2019. Her current research interests include Internet of Things, time sensitive network, and artificial intelligence.